\documentclass[apl,letter,twocolumn,epsf]{revtex4}
\usepackage{epsfig}
\usepackage{subfigure}
\usepackage{amsmath}
\usepackage{amssymb}
\usepackage{array}
\usepackage{pifont}
\usepackage{graphicx}

\def\ot{$\tau^{-1}$}
\def\igs53{In$_{0.53}$Ga$_{0.47}$As}
\def\ias52{In$_{0.52}$Al$_{0.48}$As}
\begin{document}

\title{Two-Dimensional Electron Gas in InGaAs/InAlAs Quantum Wells}

\author{E.\ Diez}
\address{\'Area de F\'{\i}sica Te\'orica, Facultad de Ciencias, Universidad de Salamanca, 37008 Salamanca, Spain}

\author{Y.\ P.\ Chen}
\address{Department of Electrical Engineering,
 Princeton University, Princeton NJ 08544, U.S.A.}

\author{S. Avesque, M.\ Hilke}
\address{Department of Physics, McGill University, Montr\'eal H3A 2T8, Canada}

 \author{E.\ Peled, D.\ Shahar}
 \address{Department of Condensed Matter Physics, The Weizmann
 Institute of Science, Rehovot 76100, Israel}

\author{J.\ M.\ Cerver\'o}
\address{F\'{\i}sica Te\'orica, Facultad de Ciencias, Universidad de Salamanca, 37008 Salamanca,
Spain}

\author{ D.\ L.\ Sivco, A.\ Y.\ Cho}
\address{Bell Laboratories, Lucent Technologies, Murray Hill NJ 07974, U.S.A.}

\date{\today}

\begin{abstract}

We designed and performed low temperature DC transport
characterization studies on two-dimensional electron gases
confined in lattice-matched
In$_{0.53}$Ga$_{0.47}$As/In$_{0.52}$Al$_{0.48}$As quantum wells
grown by molecular beam epitaxy on InP substrates. The nearly
constant mobility for samples with the setback distance larger
than 50nm and the similarity between the quantum and transport
life-time suggest that the main scattering mechanism is due to
short range scattering, such as alloy scattering, with a
scattering rate of 2.2 ps$^{-1}$. We also obtain the Fermi level
at the In$_{0.53}$Ga$_{0.47}$As/In$_{0.52}$Al$_{0.48}$As surface
to be 0.36eV above the conduction band, when fitting our
experimental densities with a Poisson-Schr\"odinger model.

\end{abstract}


\maketitle


\narrowtext

Two dimensional electron gas (2DEG) confined in \igs53\  in
lattice matched InGaAs/InAlAs/InP heterostructures and
superlattices appear in many technologically important areas
ranging from high speed electronics\cite{mishra},
optoelectronics\cite{wang,tred} and spintronics\cite{nitta,koga}.
It is also an attractive 2DEG system for the study of disorder
induced quantum phase transitions~\cite{Wei,Sahar,Wei2,pan,Hilke,
Einat}. While there have been several earlier works characterizing
electronic properties of 2DEGs in \igs53/\ias52
heterojunctions\cite{kast,walu,nakata,matsu}, \igs53/InP
heterojunctions and quantum wells (QW)\cite{weiapl,frei,grutz},
there were few systematic studies characterizing 2DEGs in
\igs53/\ias52 QWs. Since many modern
structures~\cite{mishra,wang,tred,koga} are now based on
\igs53/\ias52 QWs, such characterization is of fundamental and
technological interest.

In this letter we report the characterization of electronic
properties of 2DEG in a series of lattice matched InGaAs/InAlAs QWs
grown by molecular beam epitaxy (MBE) on InP substrate (here and
after in our paper we use abbreviations InGaAs for \igs53\ and
InAlAs for \ias52). Systematic investigations of 12 such wafers with
varying design parameters in the doping layers have yielded
important information not only about carrier mobility and
scattering, but also about how doping determines the carrier
densities, from which we were also able to determine the location of
the Fermi level at the InGaAs surface.

\begin{figure}
\includegraphics[width=.6\columnwidth,angle=-90]{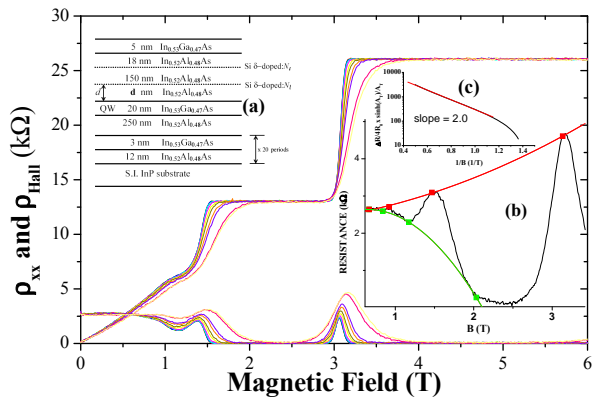}\hfill
\caption{\small The main panel shows $\rho_{xx}$ and $\rho_{Hall}$
as a function of the magnetic field for temperatures from 22 (red
trace) to 1300 mK (yellow trace), for the sample 5. The inset (a)
shows the schematic diagram of the samples. The substrate is
semi-insulating InP. The inset (b) illustrate how we extract the
amplitude of the SdH oscillations as a function of B, at 1300 mK.
Finally we plot the amplitude in a Dingle plot (c) to obtain the
scattering quantum time $\tau_q$. } \label{fig1}
\end{figure}

The schematics of the samples is depicted in Fig.~\ref{fig1}(a)
and the parameters for each sample are summarized in
Table~\ref{table1}. The 2DEG resides in a 20nm-wide InGaAs QW. Two
Si $\delta$-doped layers are placed in the InAlAs barrier to one
side (closer to the surface) of the QW. The three design
parameters that were varied are the doping densities ($N_t$ and
$N_b$) in the top and bottom dopant layers respectively, and the
distance $d$ from the bottom dopant layer to the (top) edge of the
QW.
We fabricated standard Hall bars with Indium ohmic contacts. We
tried to measure all the samples at dark. Except a few samples
($6-9$) most of them need to be illuminated to create a 2DEG. For
these samples we illuminated for sufficient time with an LED to
create a 2DEG with the highest possible mobility. We measured the
magnetoresistance $R_{xx}$ and the Hall resistance $R_{xy}$ as a
function of the perpendicular magnetic field (B) for different
temperatures. The results are shown in Fig.~\ref{fig1} for the
sample $5$ (see Table~\ref{table1}). Similar results were obtained
for the other samples. From the measured data we obtain the areal
density $n_{2D}$ and the mobility $\mu$ of the electrons, from
which we can extract the transport lifetime $\tau_t = \mu m^*/e$.
We used an effective mass $m^*$ for In$_{0.53}$Ga$_{0.47}$As of
0.043 times the bare electron mass\cite{vurgaftman}. From the
onset of SdH oscillations we extracted also the quantum lifetime
($\tau_q$) by using a Dingle style analysis. The amplitude
($\Delta R$) of the envelope function of the SdH oscillations was
found to be well described by the conventional Ando
formula~\cite{coleridge,johndavies} $\Delta R \sinh(A_T)/4 R_0 A_T
= e^{-\pi/\omega_c \tau_q}$, where $A_T=2 \pi^2 k T/\hbar
\omega_c$, $\omega_c = e B/m^*$ is the cyclotron frequency and
$R_0$ represents the resistance at zero applied magnetic field,
for a given value of temperature. To obtain the amplitudes
($\Delta R$), for each temperature we fitted the envelope of the
SdH oscillations to a pair of polynomials  as showed in
Fig.~\ref{fig1}(b) for  T=1300 mK. Next we obtained $\Delta R$
just subtracting both polynomials and we plotted $\Delta R \cdot
\sinh(A_T)/4 R_0 A_T$ versus $1/B$ in a log-x graph as in
Fig.~\ref{fig1}(c). From the previous mentioned Ando formula, we
performed a linear fit to achieve the slope ($s$) of the Dingle
plot [Fig.~\ref{fig1}(c)] and obtain the quantum lifetime for the
given temperature as $\tau_q = -(\pi \cdot m^*)/(e \cdot s)$.
Table~\ref{table1} summarizes our results measured at 4.2 K,
together with relevant parameters of the samples.
\begin{table}[t]
\caption{Sample parameters, electron densities $n_{2D}$,
mobilities $\mu$, and the transport $\tau_t$ and quantum $\tau_q$
scattering times for 12 different structures at 4.2 K.} {\small
\vspace*{0.2cm}
\begin{tabular}{c  c  c  c  c  c  c  c} \hline\hline
{\sc Sample } & $d$ & $N_t$ & $N_b$ & $n_{2D}$ & $\mu$ & $\tau_t$ & $\tau_q$\\
{\sc number} & (nm) & {\tiny($10^{11}$/cm$^{2}$)} &
{\tiny($10^{11}$/cm$^{2}$)} & {\tiny($10^{11}$/cm$^{2}$)} & {\tiny($cm^{2}/Vs$)} & (ps) & (ps) \\
\hline 1&$0$&$2$&$1$&$3.1$&$4500$&0.11&0.17\\
\hline 2&$20$&$2$&$1$&$3.0$&$13000$&0.31&0.22\\
\hline 3&$50$&$2$&$1$&$1.9$&$16000$&0.39&0.31\\
\hline 4&$50$&$1$&$1$&$1.7$&$14500$&0.35&0.31\\
\hline 5&$50$&$2$&$1$&$1.9$&$16000$&0.39&0.32\\
\hline 6&$50$&$5$&$1$&$2.2$&$15500$&0.38&0.31\\
\hline 7&$50$&$10$&$1$&$3.3$&$15000$&0.37&0.31\\
\hline 8&$50$&$10$&$0$&$2.1$&$15000$&0.37&0.31\\
\hline 9&$50$&$5$&$0.5$&$1.7$&$15000$&0.37&0.31\\
\hline 10&$150$&$2$&$1$&$1.6$&$15500$&0.38&0.33\\
\hline 11&$200$&$10$&$1$&$1.7$&$15000$&0.36&0.31\\
\hline 12&$300$&$10$&$1$&$1.5$&$15500$&0.38&0.31\\
\hline
\end{tabular}
} \label{table1}
\end{table}

\begin{figure}
\includegraphics[width=.5\columnwidth,angle=-90]{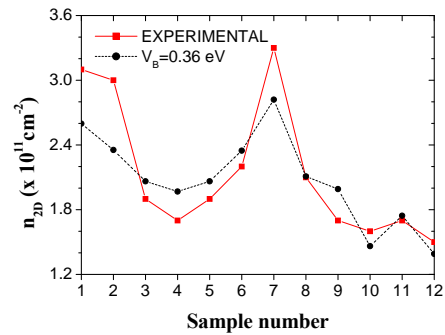}\hfill
\caption{\small The density of the 2DEG (n$_{2D}$) is plotted for
the different samples.  Lines are guides to the eye. We used
$V_B$=0.36 (eV) as the best fit to the experimental data for the
different samples.} \label{fig2}

\end{figure}
Our measurements show no indication of parallel conduction nor any
presence of a second subband. We have also confirmed this by
solving self-consistently the Schr\"odinger and Poisson equations
to calculate the subband energy levels. The calculation shows that
only one subband is occupied and the second subband is more than
40 meV above the Fermi level. To understand the origin of the
electrons forming the 2DEG, we solved analytically the
electrostatic Poisson equation for our structure following similar
procedures as in Ref.~\onlinecite{johndavies} and assuming full
ionization of dopants. This leads to the following analytical
expression for $n_{2D}$ for our structure,
\begin{equation}
n_{2D} = \frac{ N_b (c+l_1+l_2)+ N_t (c+l_2) +
(V_B\epsilon_{0}\epsilon_{B})}{c+l_1+l_2+d + z},
\label{yongeqn}
\end{equation}
where $d$, $N_t$, $N_b$ are as defined earlier, $c$=50nm is the
thickness of the InGaAs cap layer, $l_2$=18nm is the distance from
the top Si $\delta$-doped layer to the upper-end of the InAlAs and
$l_1$=150nm the distance between the two Si $\delta$-doped layers;
$\epsilon_B$=14.2 is the relative dielectric constant of
InGaAs\cite{Watling} and $V_B$ is the offset of the fermi level at
the InGaAs surface with respect to the conduction band edge. In
the denominator, $z$ is an effective ``quantum" depth of the
2DEG\cite{johndavies} which turns out to be approximately the QW
width (20nm) in our case (the deviation is negligible compared to
the contribution from other terms in the denominator).  If we
plug-in all the relevant numerical values, we get the following
dimensionless formula for $n_{2D}$:
\begin{equation}
n_{2D} = \frac{173 \, ‡ N_b \;+\; 23 \, ‡ N_t + 785 \, ‡ V_B}{d +
193}, \label{yongeq}
\end{equation}
where $n_{2D}$, $N_b$, and $N_t$ are in units of
$10^{11}\,$cm$^{-2}$, $d$ in nm, and $V_B$ in eV.  The only free
parameter in our model, $V_B$,  is the Fermi level relative to the
conduction band edge (CBE) at the InGaAs surface. Due to surface
states, the local Fermi level at a semiconductor surface is often
\text{pinned} regardless of doping and carrier density. For
example, the surface Fermi level for GaAs is about $\sim0.7\,$eV
below the CBE. For InGaAs, such information is largely unknown,
hence we have fitted this parameter using our $n_{2D}$ data and
the best fit is obtained for $V_B=0.36\,$eV. Here, the surface
Fermi level is above the CBE, which is very similar to the InAs
case~\cite{tsui70}, except that in our samples, it appears that
the surface carriers are not mobile enough and do not contribute
significantly to the transport.

\begin{figure}
\includegraphics[width=.6\columnwidth,angle=-90]{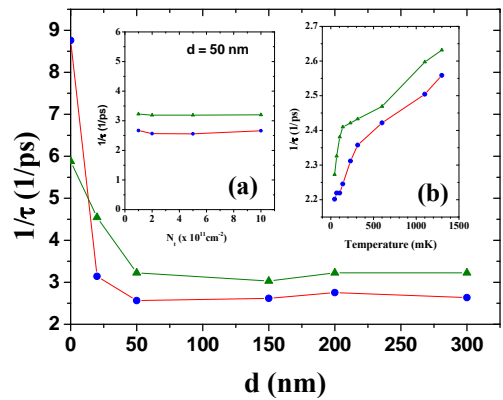}\hfill
\caption{\small The dependence of the quantum (triangles) and
transport (circles) scattering rates ($\tau^{-1}$) with the
distance $d$ of the doping layer to the quantum well . The insets
shows the dependence of $\tau^{-1}$ with (a) the amount of doping
$N_t$ and (b) temperature for the case of d=50 nm.}\label{fig3}
\end{figure}

In Fig.~\ref{fig2}, $n_{2D}$ is plotted as a function of sample
number, with $V_B=0.36$eV. Clearly, there is an excellent
agreement, between this fit (\ref{yongeq}) and the experimental
data. This also confirms that $n_{2D}$ is only a function of $d$,
$N_b$ and $N_t$ and indicates that unintentional doping from the
residual or background impurities does not appear to be
significant in our samples. Indeed, introducing background
impurities in our model would lead to a decrease of the features
seen in Fig.~\ref{fig2} and therefore does not fit our data as
well.
From the data in Table~\ref{table1}, we can notice that for $d$
larger than 50nm, the mobilities and the lifetimes are independent
of the doping parameters ($d$, $N_b$ and $N_t$). In
Fig.~\ref{fig3} we plot the quantum scattering rate
$\tau_q^{-1}$(triangles) and the transport scattering rate
$\tau_t^{-1}$ (circles), measured at 4.2K for samples 1-3 and
10-12, as a function of $d$, the distance from the bottom doping
layer to the quantum well. Both scattering rates (\ot) shows a
fast decrease at small values of $d$ (below $\sim$30nm),
indicating that the dopants provide efficient scattering for
electrons at short $d$. For $d$ of 50 nm or more, however, \ot\
becomes independent of $d$. From samples with $d$=50nm but
different $N_t$ and $N_b$, we found \ot\ to be also independent of
the doping densities, as is shown, for example, in the inset (a)
of Fig.~\ref{fig3} where \ot\ is plotted against $N_t$ at $d$=50
nm. Our findings indicate that, for $d\ge$50nm, the dopant layers
are not the major source of carrier scattering in these
structures. The common and reproducible value of mobility
($\sim$15000 cm$^{2}$/Vs) observed for wafers (with $d\ge$50 nm)
from different MBE growth suggests that the mobility, or carrier
scattering originates from some intrinsic, non-doping related
scattering in our structures. Moreover, since the dependence of
the mobility on the 2DEG density is very weak as seen in
Table~\ref{table1}, we believe that an important source of
scattering is due to the random alloy scattering potential, which
is expected to have a weak density dependence\cite{Bastard}. Such
intrinsic alloy scattering is clearly very important in the
In$_{0.53}$Ga$_{0.47}$As channel, as claimed previously to
dominate the low temperature scattering in InGaAs/InAlAs
heterojunctions\cite{walu,matsu}. We further believe that surface
roughness is less important since these structures are lattice
matched MBE grown and because surface roughness would lead to a
stronger dependence of mobility on density. Hence the main source
of disorder is short-ranged in contrast to charged doping
disorder, which is long ranged, in relation to the Fermi
wavelength. The short range nature of the dominant scattering
mechanism for our samples with $d\ge$50nm is consistent with our
observation that the quantum life time is similar to the transport
lifetime. We have measured the scattering rate dependence at lower
temperatures ($T$), as shown in inset (b) of Fig.~\ref{fig3} for
representative data in sample 5, from which we extract a low $T$
limiting value of the scattering rate to be $\sim$2.2 ps$^{-1}$.

Since most samples have to be illuminated in order to obtain the
optimum mobility, it is likely that DX centers play a
role~\cite{maude,chadi,baraldi}. We therefore estimated their
contribution by fitting, for the sample 5, the temperature
dependence of the total density to $n_{2D}=n_{free} + n_{DX}$
where $n_{free}=1.17 \times 10^{11}$cm$^{-2}$ and
$n_{DX}(T)=n^{\prime}_{DX} \exp{[(E_{DX}-E_F)/kT]}$. The best fit
is obtained for $n^{\prime}_{DX}=2.5 \times 10^{10}$cm$^{-2}$ and
$E_{DX}= E_F - 21.5$ meV (below the Fermi energy). Hence, DX
centers could explain the observed increasing of both the 2DEG
density and of the scattering rate with increasing temperature.
Indeed, at higher temperatures, more carriers are activated, which
will also leave the DX centers unsaturated and lead to the
increased scattering rate as seen in Fig.~\ref{fig3}(b).

In conclusion, we have studied the two-dimensional electron system
confined in MBE-grown
In$_{0.53}$Ga$_{0.47}$As/In$_{0.52}$Al$_{0.48}$As quantum wells.
We have measured 12 different wafers specifically designed and
grown to investigate transport properties of this type of
material. We have obtained an analytical expression for the
$n_{2D}$ of our samples with a very good agreement with the
measured values, showing that $n_{2D}$ depends only on parameters
of the Si $\delta$-doped layers ($d$, $N_t$ and $N_b$). We obtain
an excellent fit to our experimental densities assuming the
In$_{0.53}$Ga$_{0.47}$As surface Fermi energy to be $0.36\,eV$
above the conduction band. For setback distance $d$ of 50nm or
more, quantum and transport scattering rates are independent of
parameters of the dopant layers and are likely mainly due to
short-range scattering, such as alloy scattering, for which a
scattering rate of 2.2 ps$^{-1}$ is extracted at 22 mK.


We thank to Dan Tsui and Leonard Brillson for helpful discussions.
E. Diez acknowledges supports from MEC (Ram\'on y Cajal and
FIS2005-01375), JCyL (SA007B05), EC
(DIEZ-WISSMC/RITA-CT-2003-506095) and the hospitality of the Braun
Submicron Center. Y.Chen acknowledges supports from NSF and a
Gordon Wu Fellowship.

\end{document}